# Modifying the Magnetoelectric Coupling in TbMnO$_3$ by low-level Fe$^{3+}$ Substitution

A. Maia[1], R. Vilarinho[2], C. Kadlec[1], M. Lebeda[1], M. Mihalik jr.[3], M. Zentková[3], M. Mihalik[3], J. Agostinho Moreira[2], S. Kamba[1]

[1]*Institute of Physics of the Czech Academy of Sciences, Na Slovance 2, 182 21 Prague 8, Czech Republic*

[2]*IFIMUP, Physics and Astronomy Department, Faculty of Sciences, University of Porto, Rua do Campo Alegre 687, s/n- 4169-007 Porto, Portugal*

[3]*Institute of Experimental Physics of the Slovak Academy of Sciences, Watsonova 47, Košice, Slovak Republic*

**Abstract**

We report a comprehensive study of the low-level substitution of Mn$^{3+}$ by Fe$^{3+}$ effect on the static and dynamic magnetoelectric coupling in TbMn$_{1-x}$Fe$_x$O$_3$ (x=0, 0.02 and 0.04). The cationic substitution has a large impact on the balance between competitive magnetic interactions and, as a result, on the stabilization of the magnetic structures and ferroelectric phase at low temperatures. Low-lying electromagnon excitation is activated in the cycloidal modulated antiferromagnetic and ferroelectric phase in TbMnO$_3$, while it is observed up to T$_N$ in the Fe-substituted compounds, pointing for different mechanisms for static and dynamic magnetoelectric coupling. A second electrically active excitation near 40 cm$^{-1}$ is explained by means of Tb$^{3+}$ crystal-field effects. This excitation is observed up to room temperature, and exhibits a remarkable 15 cm$^{-1}$ downshift on cooling in Fe-substituted compounds. Both electromagnon and crystal-field excitations are found to be coupled to the polar phonons with frequencies up to 250 cm$^{-1}$. Raman spectroscopy reveals a spin-phonon coupling below T$_N$ in pure TbMnO$_3$, but the temperature where the coupling start to be relevant increases with Fe concentration and reaches 100 K in TbMn$_{0.96}$Fe$_{0.04}$O$_3$. The anomalies in the T-dependence of magnetic susceptibility above T$_N$ are well accounted by spin-phonon coupling and crystal-field excitation, coupled to oxygen motions.

**Corresponding authors:**
Joaquim Agostinho Moreira: jamoreir@fc.up.pt
Stanislav Kamba: kamba@fzu.cz

## 1. Introduction

Magnetoelectric multiferroics are outstanding materials as they exhibit coupled ferroelectric and magnetic orderings in the same thermodynamic phase. [1,2] The case of type-II multiferroics is more interesting, since ferroelectricity is a consequence of a particular magnetic ordering that breaks the spatial inversion symmetry of the crystal, leading to a large magnetoelectric coupling. [1,3] TbMnO$_3$ is a typical representative of type-II multiferroics. [4] This material exhibits the GdFeO$_3$-type lattice distortion, as well as, additional MnO$_6$ octahedra distortion arising from the cooperative Jahn-Teller effect associated with the Mn$^{3+}$ cation. [5] Above T$_N$ = 41 K, TbMnO$_3$ is known to be a paramagnet and paraelectric material. [4] At T$_N$, it undergoes a transition into an antiferromagnetic phase in which the Mn$^{3+}$ magnetic momenta order in the *ab*-plane with a collinear sinusoidal incommensurate modulated structure, with wavevector along the *b*-axis. [6,7] Below T$_C$ = 28 K, the spin structure transits into a cycloidal incommensurate antiferromagnetic, with the Mn$^{3+}$ spins lying in the *bc*-plane and the wavevector keeping the *b*-direction. [6,7] This magnetic structure induces ferroelectricity via the inverse Dzyaloshinskii-Moriya mechanism. [8,9] The ferroelectric polarization develops along the *c*-axis and can be rotated towards the *a*-axis when a magnetic field higher than 5.5 T is applied along the *b*-axis, as it rotates the spin cycloidal to the *ab*-plane. [4] The Tb$^{3+}$ magnetic momenta order antiferromagnetically below T'$_N$ = 7 K. [4]

The dynamic magnetoelectric coupling in TbMnO$_3$ leads to the existence of electrically active magnons, called electromagnons [10], which have been studied by inelastic neutron scattering [11–13], Raman [14,15], IR [16–18], and THz [10,19–23] spectroscopies. Electromagnons in TbMnO$_3$ are excited by the electric field component of the THz radiation, $E^\omega$, polarized parallel to the *a*-axis, regardless of the cycloidal plane orientation. [19,24] For this reason, the electromagnon activation is attributed to exchange striction, despite the static ferroelectric polarization originates from the inverse Dzyaloshinskii-Moriya mechanism. [24,25] Therefore, the static and dynamic magnetoelectric couplings in TbMnO$_3$ have different origins. However, Dzyaloshinskii-Moriya-activated electromagnons are also present, although with a much weaker dielectric strength. [23,26,27] These are observed in the $E^\omega \parallel c$ polarized spectra, after rotating the spin cycloid by applying an external magnetic field along the *b*-axis. [9]

Despite extensive literature on this material, an abnormal temperature dependence of some physical quantities in the paramagnetic and paraelectric phase has been reported, well above T$_N$, which is still matter of controversy. [28,29] The deviation of the paramagnetic susceptibility from the Curie-Weiss behavior around 200 K [28] and the anomalous temperature dependence of specific heat at this temperature [30] have been interpreted as manifestations of short-range magnetic interactions.

Moreover, a negative thermal expansion along the *a*-axis up to 150 K [29], along with anomalies in the thermal conduction coefficient at 150 K [30] and in the frequency of polar phonons assigned to both stretching and bending vibrations of $MnO_6$ octahedra [31,32], are considered a result of the crystal field effect. Recent muon spin spectroscopy (µSR) experiments rule out any short-range magnetic interaction above $T_N$. [33] It is worth stressing that a similar negative thermal expansion has been observed in $TbFeO_3$ around 200 K, evidencing that such effect should arise from only the $Tb^{3+}$ and $O^{2-}$ ions. [34]

The partial isovalent substitution of Jahn-Teller active $Mn^{3+}$ cation by the non-Jahn-Teller $Fe^{3+}$, with the same ionic radius for the 6$^{th}$ coordination, is a way to tune the symmetry-adapted lattice distortions, and therefore, the exchange interactions in the $TbMn_{1-x}Fe_xO_3$ solid solution. [35] Focusing on the low-level substitution, $x < 0.05$, the phase sequence remains nearly the same, with the $T_N$ and $T_C$ values decreasing with increasing $x$, while T'$_N$ is $x$-independent. [36] As $x$ increases from 0 to 0.04, the lowest temperature magnitude of the ferroelectric polarization decreases, which has been ascribed to the gradual fading out of the cycloidal magnetic ordering with increasing $Fe^{3+}$ content, eventually leading to the suppression of ferroelectricity for $x \geq 0.05$. [35] This hints a strong effect of the low-level substitution of $Mn^{3+}$ by $Fe^{3+}$ on the static magnetoelectric coupling. These results stimulate new experimental studies in order to unravel the effect of low-level substitution on both the dynamical magnetoelectric and elementary excitations coupling.

In this work, we report the experimental study of the dynamical magnetoelectric coupling, spin-phonon coupling and crystal-field effects on both the macroscopic and microscopic properties in oriented single crystalline $TbMn_{1-x}Fe_xO_3$ samples, with $x = 0$, 0.02 and 0.04, by means of polarized THz, IR and Raman scattering spectroscopies. Our aim is to characterize the anisotropic magnetic and polar properties of the solid solutions, to study both the static and the dynamical magnetoelectric couplings as a function of $x$, and to evidence the crystal-field effects as the main driver inducing non-symmetry breaking anomalous temperature dependence of several physical properties above $T_N$.

This paper is organized as follows. The description of the experimental details and data analysis is provided in Sec. 3. The experimental results concerning the characterization of the macroscopic magnetic and polar properties, the temperature dependence of both electromagnon, crystal field excitations and Brillouin zone center lattice vibrations are presented in Sec. 4. Finally, the discussion of the Fe-substitution effects on both macroscopic and microscopic properties is provided in Sec. 5. The paper ends with a summary of the main outcomes.

## 2. Experimental Details

Crystalline TbMn$_{1-x}$Fe$_x$O$_3$ samples with $x$ = 0, 0.02 and 0.04 were prepared by the floating zone method in air atmosphere in an FZ-T-4000 (Crystal Systems Corporation) mirror furnace. As starting materials oxides of MnO$_2$, Tb$_4$O$_7$ (both, purity 3N; supplier: Alpha Aesar) and Fe$_2$O$_3$ (purity 2N, supplier: Sigma Aldrich) were used. The starting materials were mixed in a Tb:Mn:Fe stoichiometric ratio as intended for the final compound, cold pressed into rods and sintered at 1100 °C for 12–14 h in air. [36] The single crystals were all oriented (error below 5 degrees), cut and polished, when necessary, for each measurement along each crystallographic axis.

Low-field DC induced magnetization measurements were carried out using a commercial Quantum Design superconducting quantum interference SQUID magnetometer under an applied magnetic field of 40 Oe and with a resolution better than 5x10$^{-7}$ emu.

The complex transmittance in the THz range (6 - 90 cm$^{-1}$) was measured using a custom-made time-domain spectrometer powered by a Ti:sapphire femtosecond laser with 35-fs-long pulses centered at 800 nm. The detection is performed via electro-optic sampling of the electric field of the transients within a 1-mm-thick, (110)-oriented ZnTe crystal as a sensor. [37] Low-temperature IR reflectivity measurements in the 50 – 650 cm$^{-1}$ frequency range were performed using a Bruker IFS-113v Fourier-transform IR spectrometer. The detector is a Si bolometer, cooled to 1.6 K using liquid He. For both the THz complex transmittance and IR reflectivity measurements, the temperature was controlled by Oxford Instruments Optistat optical continuous He-flow cryostats, with mylar and polyethylene windows, respectively. The IR spectra were fitted using a generalized-oscillator model with the factorized form of the complex permittivity:

$$\epsilon(\omega) = \epsilon'(\omega) + i\epsilon''(\omega) = \epsilon_\infty \prod_j \frac{\omega_{LOj}^2 - \omega^2 + i\omega\gamma_{LOj}}{\omega_{TOj}^2 - \omega^2 + i\omega\gamma_{TOj}} \qquad (1)$$

where $\omega_{TOj}$ and $\omega_{LOj}$ stand for transverse and longitudinal frequencies of the $j$-th polar phonon, respectively, and $\gamma_{TOj}$ and $\gamma_{LOj}$ the corresponding damping constants. The high-frequency permittivity, $\epsilon_\infty$, originating from electronic absorption processes, was obtained from the room-temperature frequency-independent reflectivity tail above the phonon frequencies and was assumed temperature independent. The dielectric strength, $\Delta\epsilon_j$, of the $j$-th mode is: [38]

$$\Delta\epsilon_j = \frac{\epsilon_\infty}{\omega_{TOj}^2} \frac{\prod_k \omega_{LOk}^2 - \omega_{TOj}^2}{\prod_{k \neq j} \omega_{TOk}^2 - \omega_{TOj}^2} \qquad (2)$$

The complex permittivity, $\epsilon(\omega)$, is related to the reflectivity, $R(\omega)$, by:

$$R(\omega) = \left| \frac{\sqrt{\epsilon(\omega)} - 1}{\sqrt{\epsilon(\omega)} + 1} \right|^2 \qquad (3)$$

The unpolarized Raman spectra were measured using a Renishaw inVia Qontor spectrometer with a 532 nm linearly polarized diode-pumped laser and an edge filter. The spot size of the focused laser on the sample surface is estimated to be 2 μm diameter. The temperature was controlled through a closed-cycle helium cryostat. The laser power was set at 3.3 mW to prevent heating the sample. The temperature dependence of the frequency of a given Raman mode, $\omega(T)$, were obtained through the best fit of the Raman spectra with a sum of damped oscillators: [39]

$$I(\omega, T) = [1 + n(\omega, T)]^{-1} \sum_j \frac{\omega A_{0j} \Omega_{0j}^2 \Gamma_{0j}}{\left(\Omega_{0j}^2 - \omega^2\right)^2 + \omega^2 \Gamma_{0j}^2} \qquad (4)$$

where $n(\omega, T)$ is the Bose-Einstein factor and $A_{0j}, \Omega_{0j}, \Gamma_{0j}$ are, respectively, the strength, wavenumber, and damping coefficient of the *j*-th oscillator. In the temperature range where no anomalous behavior is observed, $\omega(T)$ can be well described by the normal anharmonic temperature effect due to volume contraction as temperature decreases: [5]

$$\omega(T) = \omega_0 + C \left[1 + \frac{2}{e^x - 1}\right] \qquad (5)$$

with $x \equiv \hbar\omega_0/2k_B T$ and where $\omega_0$ and $C$ are model constants, $\hbar$ is the reduced Planck constant and $k_B$ is the Boltzmann constant.

## 3. Experimental Results

### 3.1. Temperature dependence of magnetization and electric polarization

Figure 1 shows the temperature dependence of the ratio between the applied magnetic field *H* strength and the induced magnetization *M*, $H/M$, of TbMn$_{1-x}$Fe$_x$O$_3$ (*x* = 0, 0.02, and 0.04) measured in zero-field conditions, with a magnetic field of 40 Oe applied along the *c*-axis. The magnetic phase sequence is well ascertained from the anomalies in the temperature dependence of $H/M$ below 50 K. Following reference [35], the paramagnetic - collinear sinusoidal incommensurate antiferromagnetic phase transition occurs at T$_N$ = 41 K for TbMnO$_3$, 39 K for TbMn$_{0.98}$Fe$_{0.02}$O$_3$, and 35 K for TbMn$_{0.96}$Fe$_{0.04}$O$_3$. On further cooling, the compounds exhibit another magnetic phase transition into a cycloidal modulated spin structure, which allows for ferroelectric polarization developing at T$_C$ = 28 K in TbMnO$_3$, 22 K in TbMn$_{0.98}$Fe$_{0.02}$O$_3$, and 18 K in TbMn$_{0.96}$Fe$_{0.04}$O$_3$. The lowest temperature anomalies observed in each curve in Figure 1 are assigned to the ordering of the Tb$^{3+}$ spins, occurring at T'$_N$ = 7 K for TbMn$_{0.98}$Fe$_{0.02}$O$_3$, and at 5 K for TbMn$_{0.96}$Fe$_{0.04}$O$_3$.

The Curie-Weiss law is fulfilled above 200 K, as it can be seen in the inset of Figure 1. The Curie-Weiss temperature, $\Theta_p$, determined through the best fit of the Curie-Weiss law to the $H/M$ data in the 200 – 300 K range, is found to be a decreasing function of $x$: $\Theta_p$ = -150 K for TbMnO$_3$, -165 K for TbMn$_{0.98}$Fe$_{0.02}$O$_3$, and -180 K for TbMn$_{0.96}$Fe$_{0.04}$O$_3$. The ratio $T_N/|\Theta_p|$ decreases as $x$ increases, revealing that the ferromagnetic exchange interactions between the antiferromagnetic sublattices are reinforced, meaning that the antiferromagnetic ordering becomes less stable as $x$ increases. Table I summarizes the critical temperatures, ascertained from the temperature anomalous dependence of H/M, and the $\Theta_p$ values.

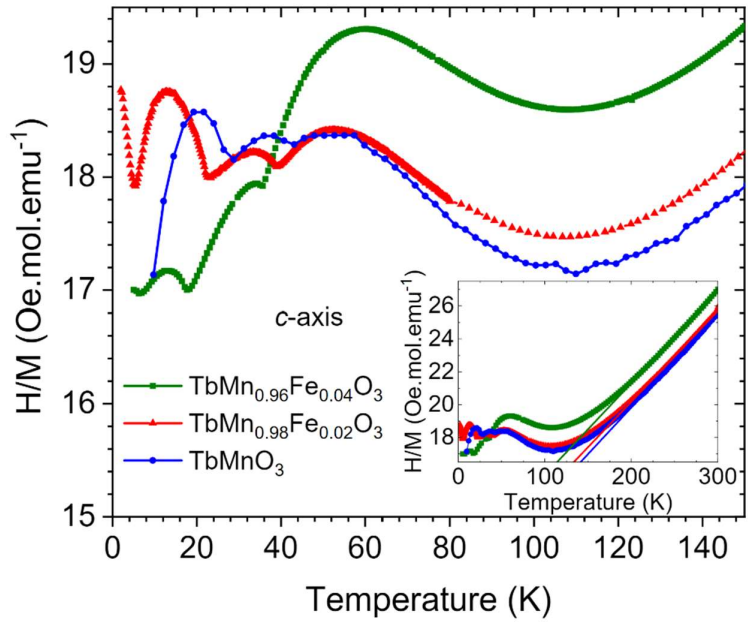

**Figure 1.** Temperature dependence of the $H/M$ ratio, measured in ZFC conditions, with a $0 Oe strength magnetic field applied along the c-axis, for TbMnO$_3$, TbMn$_{0.98}$Fe$_{0.02}$O$_3$ and TbMn$_{0.96}$Fe$_{0.04}$O$_3$. Inset: $H/M$ in the 0 – 300 K range. The dashed lines were determined from the best fit of the Curie-Weiss law above 200 K and extrapolated to intercept the temperature axis. The data for TbMnO$_3$ was obtained from reference [28].

**Table I.** Critical temperatures and Curie-Weiss temperature, $\Theta_p$, of TbMn$_{1-x}$Fe$_x$O$_3$ (0 ≤ $x$ ≤ 0.04) ascertained from the temperature dependence of the c-component of magnetization. The data for TbMnO$_3$ were obtained from reference [28].

| $x$ | T$_N$ [K] | T$_C$ [K] | T'$_N$ [K] | $\Theta_p$ [K] |
|---|---|---|---|---|
| 0.00 | 41 | 28 | 7 | -150 |
| 0.02 | 39 | 22 | 7 | -165 |
| 0.04 | 35 | 18 | 5 | -180 |

Between $T_N$ and 200 K, the $H/M = \chi^{-1}(T)$ data depart from the expected Curie-Weiss law observed above 200 K. Moreover, a broad minimum of H/M(T) is observed at around 110 K. These abnormal temperature dependences are *x*-independent, in agreement with previous reports for these compounds. [40,41] We will discuss the mechanism underlying this temperature behavior later on. Figure 2 shows the temperature dependence of the electric polarization of $TbMn_{0.98}Fe_{0.02}O_3$ and $TbMn_{0.96}Fe_{0.04}O_3$, measured along the *a*- and *c*-axes. The emergence of the electric polarization occurs at 23 K and at 18 K in $TbMn_{0.98}Fe_{0.02}O_3$ and $TbMn_{0.96}Fe_{0.04}O_3$, respectively, and these values agree with the corresponding $T_C$'s ascertained form the anomalies in the $H/M$(T) curves (see Table I). Below $T_C$, the value of the electric polarization along the *a*- or *b*-axis is about 9% of the corresponding values along the *c*-axis. Assuming that the magnetic ordering in these compounds is similar to that of $TbMnO_3$ in the ferroelectric phase, the values of the electric polarization measured along the *a*- and *b*-axes most likely originate from leakage from the *c*-axis, due to slight orientation misalignments of 5° at most.

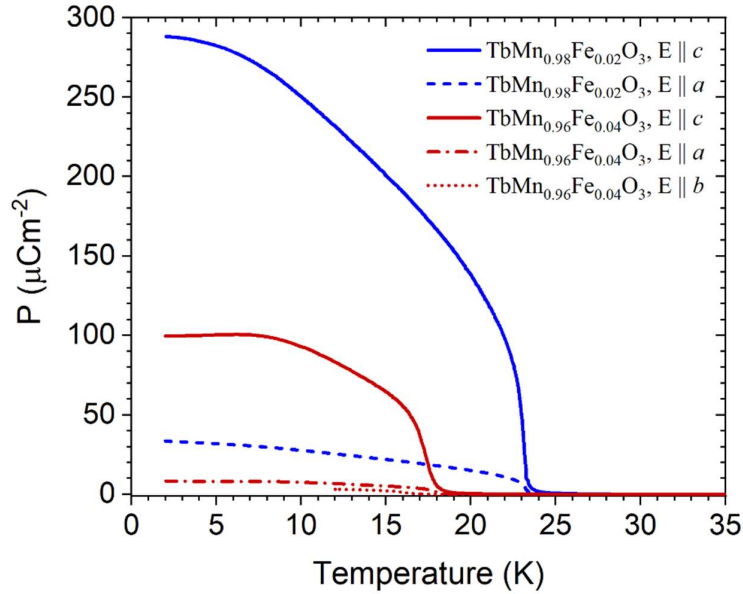

**Figure 2.** Temperature dependence of the electric polarization of $TbMn_{0.98}Fe_{0.02}O_3$ and $TbMn_{0.96}Fe_{0.04}O_3$ measured along the *a-, b-* and *c*-axes.

As already observed in ceramic samples [35], the ferroelectric phase transition temperature, $T_C$, and the electric polarization magnitude, |***P***|, decrease with increasing Fe-content up to *x* = 0.04.

## 3.2. Time-domain THz spectroscopy

Figure 3 shows, respectively, the polarized THz spectra for $E^\omega \parallel a$ and $H^\omega \parallel c$ of TbMnO$_3$, and for $E^\omega \parallel a$ and $H^\omega \parallel b$ of TbMn$_{0.98}$Fe$_{0.02}$O$_3$ and TbMn$_{0.96}$Fe$_{0.04}$O$_3$, recorded at several fixed temperatures in the 4 – 300 K range. The THz spectra of TbMn$_{0.96}$Fe$_{0.04}$O$_3$ for all possible $E^\omega$ and $H^\omega$ polarization configurations of the THz radiation are shown in Figure S1 of Supplemental Information. [42] The absorption at THz frequencies is strongest in the $E^\omega \parallel a$ polarization, independently of the $H^\omega$ polarization, as it can be concluded from the results shown in Figure S1. This result agrees with the one reported for TbMnO$_3$ [16,24] and other orthorhombic rare-earth manganites. [22,26]

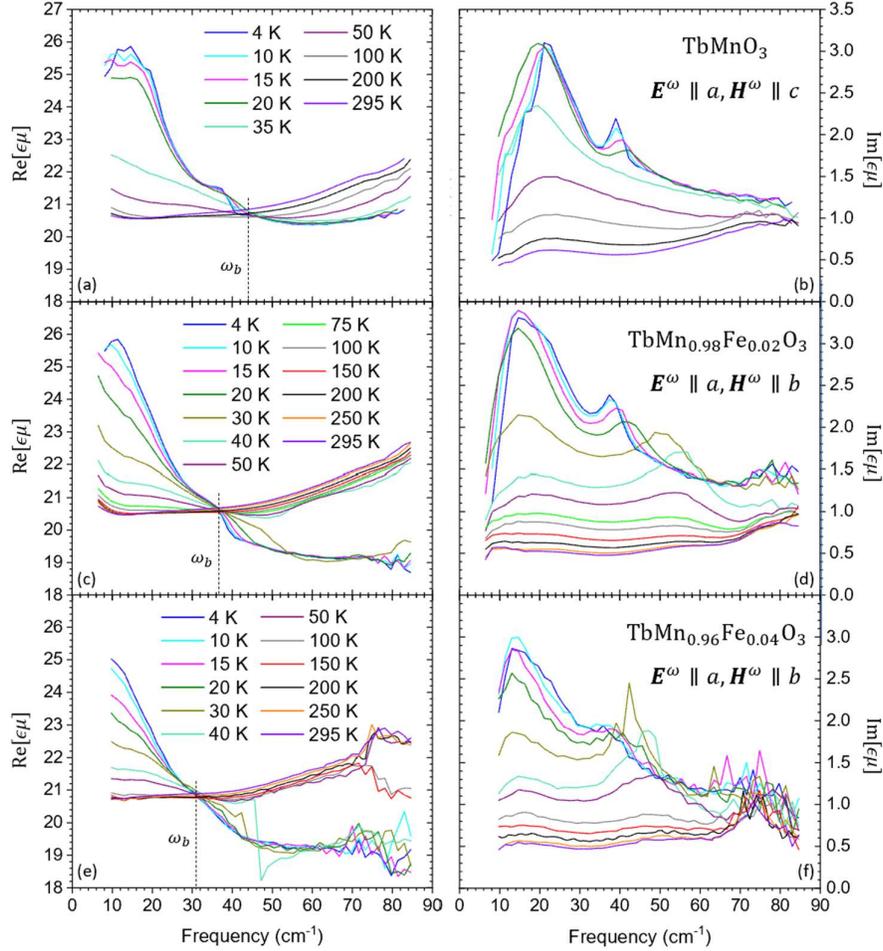

**Figure 3.** $Re[\epsilon\mu]$ and $Im[\epsilon\mu]$ THz spectra for $E^\omega \parallel a$, $H^\omega \parallel c$ of TbMnO$_3$, and for $E^\omega \parallel a$, $H^\omega \parallel b$ of TbMn$_{0.98}$Fe$_{0.02}$O$_3$ and TbMn$_{0.96}$Fe$_{0.04}$O$_3$, measured at several temperatures.

The $Re[\epsilon\mu]$ spectra can be divided into two main spectral ranges by a specific frequency, $\omega_b$ (see Figures 3(a), 3(c) and 3(e)), which decreases from 44 cm$^{-1}$ for TbMnO$_3$, to 31 cm$^{-1}$ in TbMn$_{0.96}$Fe$_{0.04}$O$_3$. Below 50 K, few degrees above T$_N$, $Re[\epsilon\mu]$ strongly increases for $\omega < \omega_b$, while decreases for $\omega >$ $\omega_b$. Such behavior evidences an oscillator strength transfer mechanism from higher frequency excitations to excitations in the $\omega < \omega_b$ range.

In diamagnetic paraelectrics, the dielectric loss, $\epsilon''(\omega)$, should be, in general, linearly dependent on frequency ($\epsilon''(\omega) \propto \omega$) far below phonon frequencies and $\lim_{\omega \to 0} \epsilon''(\omega) = 0$. [43] The absorption background observed in Figures 3(b), 3(d) and 3(f), is already seen in the loss spectra at room temperature as one (in case of TbMnO$_3$, located at 22 cm$^{-1}$) or two (in case of TbMn$_{0.98}$Fe$_{0.02}$O$_3$ and TbMn$_{0.96}$Fe$_{0.04}$O$_3$, located at around 18 - 20 cm$^{-1}$ and 50 – 55 cm$^{-1}$, respectively) broad and weak bands, whose intensity continuously increase on cooling. The absorption band in TbMnO$_3$ above T$_C$ has been ascribed to a Deby-like relaxation.[20,21] The shape of the two bands in the THz spectra of the other two compounds cannot be described by an oscillator formula neither by a Debye relaxation model. Instead, the absorption evidences more overdamped excitations. In TbMnO$_3$, the absorption background rapidly increases on cooling below 50 K, and develops in two absorption bands, peaking near 20 cm$^{-1}$ and 40 cm$^{-1}$, respectively, below T$_C$. These bands are only observed in the $\boldsymbol{E^\omega} \parallel a$ polarized spectra, independently on magnetic $\boldsymbol{H^\omega}$ orientation. [10,16,19] Similar temperature evolution is seen in the loss spectra of TbMn$_{1-x}$Fe$_x$O$_3$, $x$ = 0.02 and 0.04. The main difference is that the higher frequency band is present at all temperatures and its frequency exhibits larger temperature variation. According to references [25,44], the lowest frequency excitation, seen near 20 cm$^{-1}$, is assigned to an electrically active Brillouin zone boundary magnon (electromagnon), with wave vector $q=2\pi/b$. [7,10] The results clearly evidence that the electromagnon persists in the Fe-substituted compounds, but with different properties relatively to that detected in TbMnO$_3$, as it is already observed below T$_N$. The electromagnon character of this excitation is noticeable in the Re[$\epsilon\mu$] spectra, which exhibit an increase at lower frequencies and decrease at higher frequencies, evidencing the dielectric strength transfer from polar phonons above 80 cm$^{-1}$ to the electromagnon. These phonons are observed in the IR spectra and will be addressed later.

The highest frequency excitation needs a more detailed analysis. This excitation is observed in the paramagnetic phase, where no magnons are expected. Paraelectromagnons can exist in systems with short-range magnetic order, [45] but recent muon spin relaxation studies (μSR) in TbMn$_{1-x}$Fe$_x$O$_3$, $x$ = 0 and 0.02, and neutron scattering experiments on $x$ = 0.02 do not reveal any short-range magnetic correlations above T$_N$. [33] A more plausible explanation for the absorption near 40 cm$^{-1}$ is the crystal field excitation of the $F_6$ electron levels in Tb$^{3+}$. This crystal field excitation was observed in inelastic neutron scattering [11,12,46] and in middle IR studies [40]. Moreover, magnons seen in inelastic neutron scattering pattern recorded at 17 K have frequencies of 50 and 65 cm$^{-1}$ at $q=2\pi/b$ [11] (i.e., wavevector of the predicted electromagnon), excluding the hypothesis that this band, seen near 40 cm$^-$ [1], originates from an electromagnon. While this excitation is weakly frequency dependent in TbMnO$_3$, its frequency decreases from 55 to 40 cm$^{-1}$ on cooling from 50 to 4 K, giving evidence for a gradual change of crystalline environment of the Tb$^{3+}$ cations. This change is also seen in the anomalous temperature dependence of lattice parameters. [29,30]

The $E^\omega \perp a$ polarized $Re[\epsilon\mu]$ spectra recorded at several temperatures, shown in Figure 4, display a different temperature dependence. No spectral weight transfer is observed, unlike for $E^\omega \parallel a$. $Re[\epsilon\mu]$ decreases on cooling, specially below 60 cm$^{-1}$, but with different temperature rates which increases with increasing $x$. For TbMn$_{0.96}$Fe$_{0.04}$O$_3$, the $Re[\epsilon\mu]$ spectra exhibit a low frequency band (peaking at around 12 cm$^{-1}$ at 295 K), whose intensity decreases on cooling and eventually disappears below 75 K. The $Im[\epsilon\mu]$ spectra are nearly temperature independent above 100 K and below 50 K (see Figure S2), respectively, while in the 50–100 K range, a steep drop is observed, becoming more pronounced with increasing $x$ (see Figure S2).

An overdamped band, already observed at room temperature, develops near 18 cm$^{-1}$ in the $E^\omega \perp a$ polarized $Im[\epsilon\mu]$ spectra. Such band was reported in the paramagnetic phase of TbMnO$_3$ and other orthorhombic rare-earth manganites, having been assigned to a Debye-like relaxation. [21] Interestingly, the temperature evolution of this band is $x$-dependent. In TbMnO$_3$, on cooling, the band narrows and its intensity increases below 30 K; i.e, in the ferroelectric phase. The origin of this band in the $E^\omega \parallel c$ absorption spectra of TbMnO$_3$ (see Figure 4(b)) is ascribed to a leakage from $E^\omega \parallel a$ spectra due to a small misorientation during the sample preparation. However, this band is very strong in the absorption spectra of TbMn$_{0.96}$Fe$_{0.04}$O$_3$, with an amplitude of the same order of magnitude as the low-frequency electromagnon band observed in the $E^\omega \parallel a$ polarization spectra (see Figure 3(f)), which excludes any leakage effect due to the sample misorientation. Moreover, another overdamped band develops at 52 cm$^{-1}$ at room temperature and fades out on cooling, eventually disappearing below 50 K. This result is puzzling, because the disappearance of this second excitation occurs in the magnetically ordered phases. Interestingly, this bands becomes more intense as Fe-content increases. Further studies are needed to understand the origin of this band.

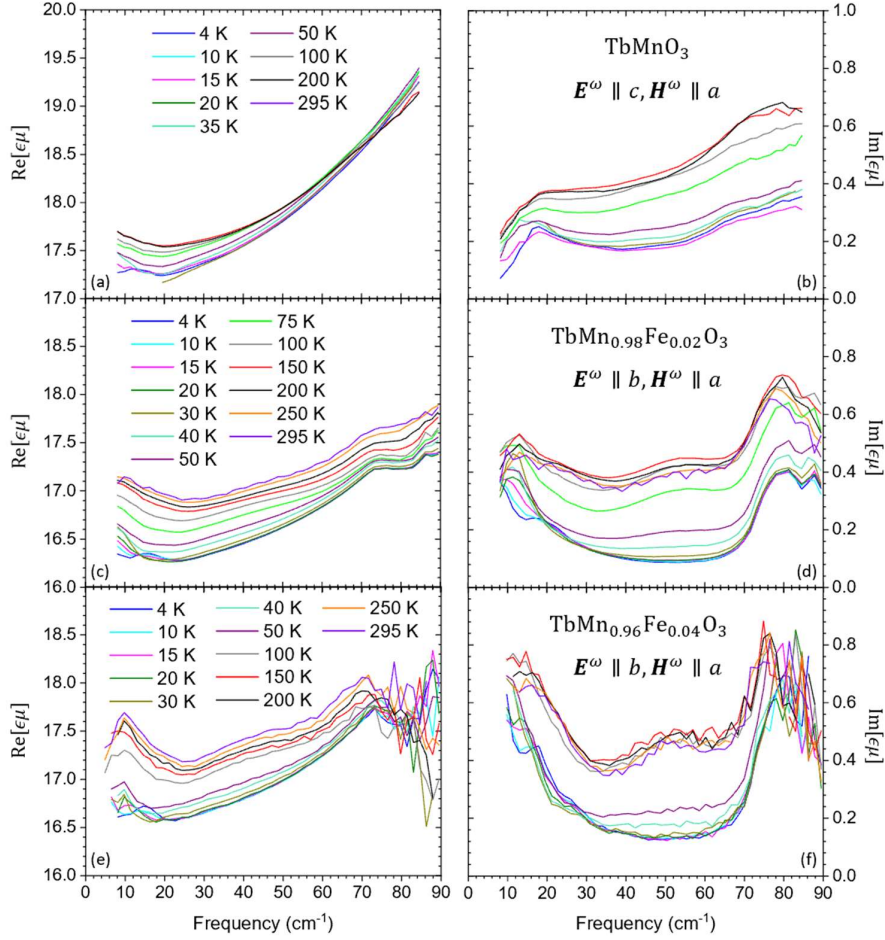

**Figure 4.** Temperature dependence of the $Re[\epsilon\mu]$ and $Im[\epsilon\mu]$ spectra of TbMnO$_3$ ((a), (b)), TbMn$_{0.98}$Fe$_{0.02}$O$_3$ ((c), (d)) and TbMn$_{0.96}$Fe$_{0.04}$O$_3$ ((e), (f)). The spectra of TbMnO$_3$ were measured for the $E^\omega \parallel c, H^\omega \parallel a$ polarization, while those of TbMn$_{0.98}$Fe$_{0.02}$O$_3$ and TbMn$_{0.96}$Fe$_{0.04}$O$_3$ correspond to the $E^\omega \parallel b, H^\omega \parallel a$ polarization.

### 3.3. Γ-point optical phonons

At room conditions, TbMnO$_3$ crystallizes in the orthorhombic $P$bnm structure. The irreducible representation decomposition for the Brillouin zone center phonons is: $\Gamma_{phon} = 7A_g \oplus 8A_u \oplus 5B_{1g} \oplus 10B_{1u} \oplus 7B_{2g} \oplus 8B_{2u} \oplus 5B_{3g} \oplus 10B_{3u}$. The $9B_{1u}, 9B_{3u}$ and $7B_{2u}$ modes are IR active for the electric field $E^\omega$ of IR radiation parallel to the $a$-, $b$-, and $c$-axes, respectively, while the $7A_g$, $5B_{1g}$, $7B_{2g}$, and $5B_{3g}$ modes are Raman active and the remaining $8A_u$ modes are silent. [18] The low-level substitution of Mn by Fe locally breaks the symmetry while keeping the same overall $P$bnm space group for the mean crystallographic structure. Therefore, we shall assume the same decomposition of the irreducible representations of the Γ-point phonons for the Fe-substituted compounds.

### 3.3.1. Polar phonons in TbMn$_{0.98}$Fe$_{0.02}$O$_3$

In this sub-section, we present the B$_{1u}$ symmetry ($\boldsymbol{E}^{\omega} \parallel a$ polarized) infrared spectra of TbMn$_{0.98}$Fe$_{0.02}$O$_3$. We have chosen this polarization as electromagnons are observed in THz spectra with the same polarization; consequently, the oscillator strength transfer from polar optical phonons to low frequency excitations can be better studied. Since the $\boldsymbol{E}^{\omega} \parallel a$ polarized THz spectra is independent of $\boldsymbol{H}^{\omega}$ polarization, the excitations in that frequency range contribute to permittivity and not to permeability. Therefore, in this section, we present the real ($\epsilon'$) and imaginary ($\epsilon''$) parts of the complex dielectric spectra without considering the contribution of $\mu(\omega)$ to the IR signature.
Representative $\boldsymbol{E}^{\omega} \parallel a$ polarized reflectivity spectra of TbMn$_{0.98}$Fe$_{0.02}$O$_3$ recorded at 4 and 250 K, are shown in Figure 5(a). The reflectivity spectra recorded in TbMn$_{0.98}$Mn$_{0.02}$O$_3$ resembles very well the ones observed for TbMnO$_3$, evidencing the similar mean crystallographic structure for both compounds. [17,18] The plotted reflectivity below 85 cm$^{-1}$ was calculated from the measured complex permittivity data in the THz range, through Equation 3, while the higher-frequency IR spectra were directly measured. Above 85 cm$^{-1}$, nine main reflectivity bands are observed, as expected from group theory, corresponding to the $B_{1u}$-symmetry phonons. However, four additional weak modes are needed to improve the quality of the fits; these additional modes are due to $B_{2u}$- and $B_{3u}$-symmetry phonons that appear in the spectra due to a slight misorientation of the sample. Two additional spectral components in the 125 -155 cm$^{-1}$ range, below 100 K, are assigned to crystal-field excitations. [40] The spectra of the real ($\epsilon'$) and imaginary ($\epsilon''$) parts of the complex permittivity, obtained from the fits of Equation 1 to the IR reflectivity spectra, are shown in Figures 5(b) and 5(c), respectively. The peaks in $\epsilon''(\omega)$ correspond to the transverse optical phonons, at frequencies $\omega_{TO}$, or to other polar excitations (e.g., crystal-field excitations or electromagnons).

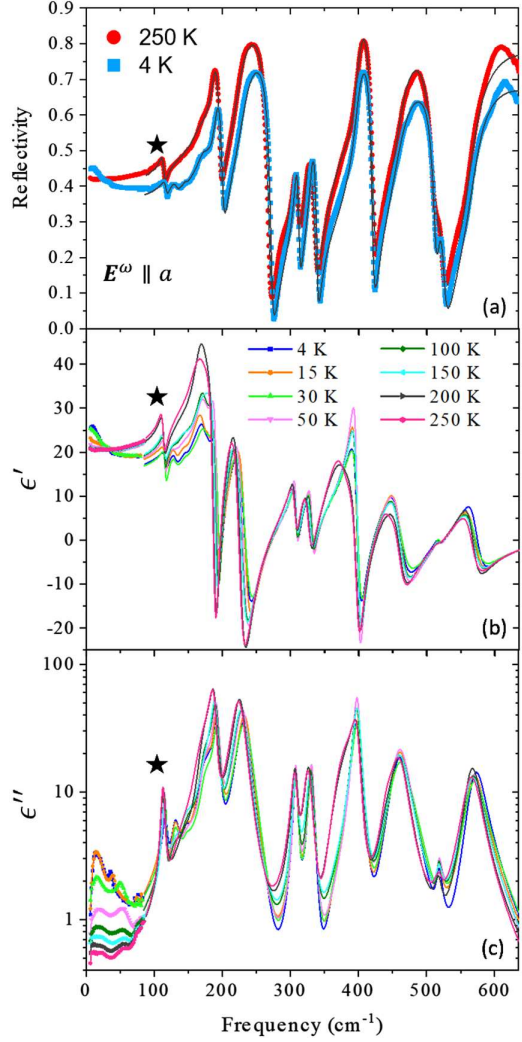

**Figure 5.** (a) IR reflectivity spectra of TbMn$_{0.98}$Fe$_{0.02}$O$_3$ recorded at 4 K and 250 K, for $E^{\omega} \parallel a$. The reflectivity below 85 cm$^{-1}$ was calculated from the THz spectra. The corresponding fits are shown as dotted lines. (b) $\epsilon'$ and (c) $\epsilon''$ spectra of TbMn$_{0.98}$Fe$_{0.02}$O$_3$ obtained from fitting the IR reflectivity at several temperatures. The low-frequency data seen below 85 cm$^{-1}$ corresponds to the experimental THz spectra. The polar phonon around 115 cm$^{-1}$ is indicated by a star.

Figure 6 shows the contour plot of the $E^{\omega} \parallel a$ polarized $\epsilon''(\omega)$ spectra of TbMn$_{0.98}$Mn$_{0.02}$O$_3$ which better evidences the temperature evolution of the polar excitations. On cooling from 300 K, several transversal optical phonons exhibit interesting changes. The collinear sinusoidal modulated antiferromagnetic magnetic phase transition at T$_N$ is well signaled by the intensity decrease (about one order of magnitude) of the polar phonons observed in the 160 – 250 cm$^{-1}$ and 350 - 400 cm$^{-1}$ spectral ranges. Interesting, the phonon at around 400 cm$^{-1}$ recovers intensity below T$_C$ = 18 K. A 50 % dielectric strength reduction below T$_N$ has been also reported for the lowest frequency phonon in TbMnO$_3$ (~114 cm$^{-1}$ at 300 K), while the phonon at 404 cm$^{-1}$ reduces its oscillator strength on cooling towards T$_N$, where the minimum value is reached, increasing its value on further cooling. [16] Although not clear evident in Figure 6, the polar phonon around 115 cm$^{-1}$ also decreases its dielectric strength on cooling below 100 K (see Figure 5(c)). These results give evidence for the dielectric strength transfer below 100 K from the polar phonons observed up to 250 cm$^{-1}$ to the electromagnon and crystal-field excitations in TbMn$_{0.98}$Mn$_{0.02}$O$_3$. This evidences the coupling of all these excitations, which exhibit change of strength with temperature. Unfortunately, no lattice dynamic calculations of polar phonons in TbMnO$_3$ are reported to the best of our knowledge. So, considering their frequencies, we tentatively assignment the bands showing decrease of dielectric strength to vibrations involving the oxygen atoms and the Mn/Fe cations. These results point towards distortions of the crystal lattice in the magnetic phase transitions, likely involving oxygen shifts.

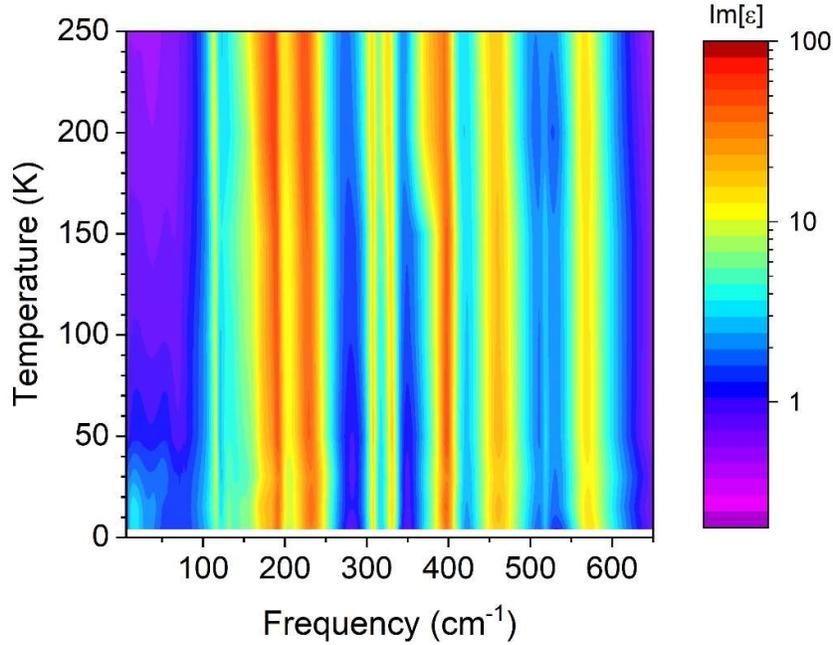

**Figure 6.** Temperature evolution of the $\epsilon''$ spectra of TbMn$_{0.98}$Fe$_{0.02}$O$_3$ for $\boldsymbol{E^\omega} \parallel a$.

Interesting changes in the IR spectra are also evident in the paramagnetic phase of TbMn$_{0.98}$Mn$_{0.02}$O$_3$. These are best observed in the 250 – 500 cm$^{-1}$ range, where polar phonons involving vibrations of the oxygen atoms are expected. The main changes observed concern the oscillator strength changes; modes at 280 cm$^{-1}$ and 350 cm$^{-1}$ decrease their oscillator strength below 175 K, while that of the mode at 398 cm$^{-1}$ increases below this temperature. It is worth to stress that it is precisely this last mode that loses strength on entering in the collinear sinusoidal incommensurate modulated antiferromagnetic phase. The aforementioned results evidence oxygen octahedra distortions well above $T_N$, which is a well-known feature of TbMnO$_3$, and therefore, independent on Fe-substitution.

### 3.3.2 Raman modes

Representative unpolarized Raman spectra of TbMn$_{1-x}$Fe$_x$O$_3$, $x = 0$, 0.02 and 0.04, recorded at 10 K in the 100 – 800 cm$^{-1}$ spectral range, are shown in Figure 7. The unpolarized Raman spectra recorded at several temperatures in the 10 – 300 K range, are found in Figures S3, S4 and S5 (see Supplemental Information). [42] The Raman signature of the three compositions here studied is rather similar, as a consequence of the similar structure of these compounds, as previously evidenced. The Raman mode assignment for TbMnO$_3$ has been published by various groups, [47–49] and can be summarized as follows: the Raman-active phonons below 200 cm$^{-1}$ are mainly characterized by vibrations of the heavy rare-earth ions and, above 300 cm$^{-1}$, by motions of the oxygen ions. The B-site atoms do not contribute for the phonons observed by Raman light scattering. The activation of new Raman bands is not observed in the ferroelectric phase transition at $T_C$, as it is common in magnetically-driven ferroelectric materials. However, a detailed quantitative analysis of the Raman spectra reveals interesting temperature dependences of some phonon frequencies.

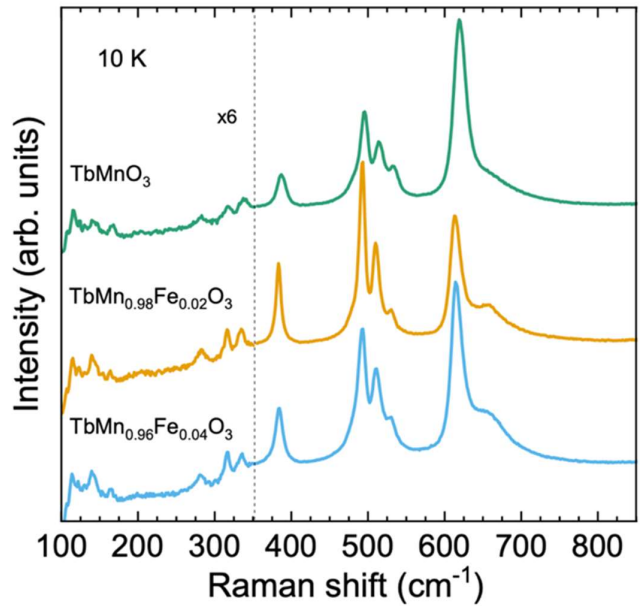

**Figure 7.** Unpolarized Raman spectra of TbMnO$_3$, TbMn$_{0.98}$Fe$_{0.02}$O$_3$ and TbMn$_{0.96}$Fe$_{0.04}$O$_3$ recorded at 10 K. The spectra below 354 cm$^{-1}$ are multiplied by 6 for clarity.

Figure 8 shows the temperature dependence of some selected Raman-active phonon frequencies in the three compositions here studied, which better mirror the structural deformations induced by temperature and the effect of the Fe-substitution on their temperature dependence. The solid lines presented in Figure 8 were determined from the best fit of Equation 5 to the experimental data above 120 K and extrapolated to low temperatures. The low frequency Raman bands at ~141 cm$^{-1}$ and ~166 cm$^{-1}$ are assigned to Tb-oscillations along the *a*- and *c*-axes, respectively. [47–49] These phonons do not display anomalous temperature dependence in TbMnO$_3$ in all temperature range studied. On contrarily, both MnO$_6$ octahedra rotation (located near 379 cm$^{-1}$) and symmetrical stretching (located near 611 cm$^{-1}$) modes in TbMnO$_3$ show an upward deviation relatively to the anharmonic temperature behavior below T$_N$. In TbMn$_{0.98}$Fe$_{0.02}$O$_3$ and TbMn$_{0.96}$Fe$_{0.04}$O$_3$, a clear downward deviation from the anharmonic temperature dependence is ascertained for the Tb(*a*) oscillations, just below 75 K and 100 K, respectively. Moreover, the BO$_6$ octahedra rotation mode follows the normal temperature dependence in all temperature range, while the symmetrical stretching mode shows a downward deviation below 75 K. The BO$_6$ symmetrical stretching mode is known to be sensitive to the magnetic exchange integrals, as it involves changes in the B−O2 bond lengths. The upwards/downward deviation demonstrates the relative balance between the ferromagnetic and antiferromagnetic interactions. [50,51] Regarding TbMnO$_3$, the upward frequency shift has been interpreted as a manifestation of the prevalence of the antiferromagnetic interactions over the ferromagnetic ones. [50,51] Following the model, in Fe-substituted compounds the downshift observed is a consequence of the reinforcement of the ferromagnetic interactions against the antiferromagnetic ones.

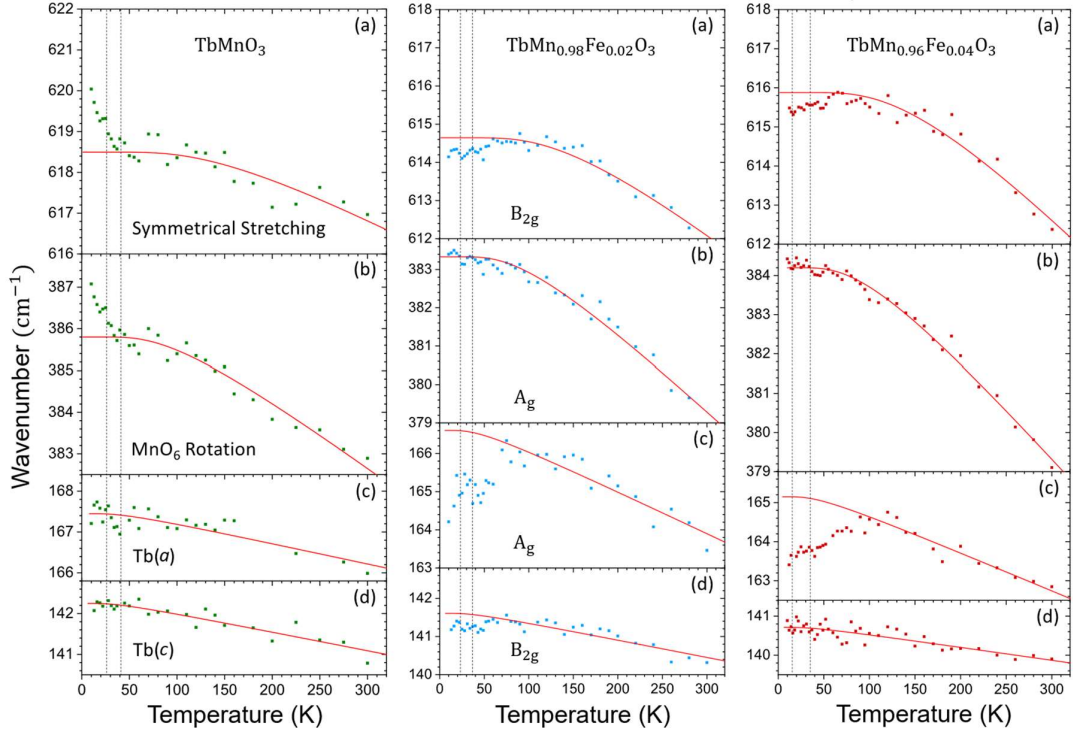

**Figure 8.** Temperature dependence of the Raman mode frequencies for TbMnO$_3$ (left panel), TbMn$_{0.98}$Fe$_{0.02}$O$_3$ (middle panel) and TbMn$_{0.96}$Fe$_{0.04}$O$_3$ (right panel). The best fit with Equation 5 in the temperature range where no anomalous behavior is observed is represented by a solid line and is extrapolated down to 0 K. Tb($a$) and Tb($c$) denote the Raman Tb$^{3+}$ oscillations along the $a$ and $c$ axis, respectively. In each panel, the dashed line at the lowest temperature marks $T_C$ while the other marks $T_N$.

## 4. Discussion

### 4.1. Reinforcement of the ferromagnetic interactions by Fe-substitution

One of the outcomes of this work concerns the high sensitivity of the magnetic and polar orderings and their stability temperature range in TbMnO$_3$ due to partial substitution of Mn$^{3+}$ by Fe$^{3+}$. We have found that the low level isovalent substitution has a large impact on the balance between competitive ferro- and antiferromagnetic interactions in TbMn$_{1-x}$Fe$_x$O$_3$. The 4% level substitution of Mn$^{3+}$ by Fe$^{3+}$ results in the $T_N$ decrease of 14%, and $T_C$ of 48%. Considering that TbMn$_{0.95}$Fe$_{0.05}$O$_3$ does not exhibit ferroelectric phase but a weak-ferromagnetic behavior, [35] the results here reported give clear evidence for the reinforcement of the ferromagnetic interactions against the antiferromagnetic ones, and the consequent destabilization of the antiferromagnetic orders with increasing $x$ up to 0.04. This interpretation is also corroborated by the decreasing $T_N/|\Theta_p|$ ratio with increasing Fe-content, and the crossover between up to down shifts observed in the BO$_6$ symmetrical stretching Raman-active

phonons, as it was pointed before. The decrease of the maximum electric polarization at lowest temperature as $x$ increases, is also an indirect proof of the spin structure changes induced by $Fe^{3+}$ in the crystal lattice. Indeed, according to the Dzyaloshinskii-Moriya model, the polarization magnitude decrease is a direct consequence of less tilted neighbor spins in the cycloidal modulated antiferromagnetic phase. The high sensitivity of the electric polarization to the low substitution level results from the rather large static magnetoelectric coupling in these materials.

**4.2. Magnetoelectric and crystal-field excitations in TbMn$_{1-x}$Fe$_x$O$_3$, $0 \leq x \leq 0.04$**

Low level Fe-substitution also has relevant effects on the dynamic magnetoelectric excitations in TbMn$_{1-x}$Fe$_x$O$_3$, $0 \leq x \leq 0.04$. The results corroborate that the magnetoelectric excitations are strictly tied to crystal lattice, although the substitution of $Mn^{3+}$ by $Fe^{3+}$ alter their activity and temperature dependence.

Clear evidence for a dynamic magnetoelectric contribution to the THz spectra is found for the three compounds, but with different properties. In TbMnO$_3$, the low energy electromagnon band peaks at 20 cm$^{-1}$ and it is observed in the $\boldsymbol{E^\omega} \parallel \boldsymbol{a}$ polarized THz spectra, in agreement with the selection rule for exchange-striction activated electromagnons. [24] Its intensity is large in the cycloidal modulated antiferromagnetic phase, which is also ferroelectric and, as temperature approaches T$_C$ from below, the electromagnon band intensity monotonously decreases and fades out above T$_C$. The THz signal observed above T$_C$ exhibits a rather overdamped excitation that has been described as a Debye-like relaxation. [21] The electromagnon band is also observed around 20 cm$^{-1}$ in the $\boldsymbol{E^\omega} \parallel \boldsymbol{a}$ polarized THz spectra of TbMn$_{1-x}$Fe$_x$O$_3$, $x$ = 0.02 and 0.04, but in both the cycloidal modulated antiferromagnetic and ferroelectric phase, and the collinear sinusoidal modulated antiferromagnetic and paraelectric phase. The electromagnon band disappears above T$_N$, remaining two broad and weak absorption bands up to 300 K, instead one as is observed in TbMnO$_3$. The existence of the electromagnon band below T$_N$ in TbMn$_{0.98}$Fe$_{0.02}$O$_3$ and TbMn$_{0.96}$Fe$_{0.04}$O$_3$ compounds gives clear evidence for the different origin of this excitation in the Fe-substituted compounds. According to the current models, the electromagnon in TbMnO$_3$ arises in the dielectric spectra due to exchange-striction mechanism ($\propto \vec{S}_i \cdot \vec{S}_j$). [24,25] Exchange-striction-activated electromagnons have been observed even in the paraelectric phases of Ba$_2$Mg$_2$Fe$_{12}$O$_{22}$ [52] and CuFe$_{1-x}$Ga$_x$O$_2$ [53], which exhibit proper screw and collinear magnetic ordering, respectively. Therefore, the activation of electromagnons is not necessarily tied to spin-induced ferroelectric phases. The results here presented for the Fe-substituted TbMnO$_3$ compounds clearly evidence the large impact that low-level Mn-substitution by $Fe^{3+}$ has in the magnetic structure and, consequently, in the mechanisms underlying the electromagnon activation.

The observation of the overdamped component in the THz spectra of $TbMnO_3$, at 20 cm$^{-1}$, and of $TbMn_{0.98}Fe_{0.02}O_3$ and $TbMn_{0.96}Fe_{0.04}O_3$, at around 18 - 20 cm$^{-1}$ and 50 – 55 cm$^{-1}$, above $T_N$ should not have magnetic origin. In fact, recent temperature and transverse magnetic field µSR studies in $TbMnO_3$, [33] performed by the authors, excludes any short (and long) range magnetic ordering above $T_N$, in agreement with inelastic neutron scattering experiments. [11,12]

The second polar absorption component, developing in the THz spectra of $TbMnO_3$ below $T_C$, near 40 cm$^{-1}$, has been tentatively assigned to a Brillouin zone boundary electromagnon, with wavevector $q = 2\pi/b - 2q_m$. [12,24,25] Nevertheless, inelastic neutron scattering studies in $TbMnO_3$ revealed magnons with higher frequencies at this wavevector. [11] In fact, Takahashi *et al* [16] reported a second electromagnon excitation near 60 cm$^{-1}$ in $TbMnO_3$. The excitation we detect in this work is 20 cm$^{-1}$ lower, which excludes its assignment to an electromagnon. Following S. Mansouri *et al*, [40] $q$-independent crystal field excitations have been observed in $TbMnO_3$, one of them having rather similar frequency as to this second broad component, which enable us to assign this component to a crystal-field excitation. Similar component is also observed in the THz spectra of $TbMn_{0.98}Fe_{0.02}O_3$ and $TbMn_{0.96}Fe_{0.04}O_3$, developing around 55 cm$^{-1}$ and 50 cm$^{-1}$, respectively, below 100 K. Interestingly, the crystal field excitation reveals a 15 cm$^{-1}$ downshift on cooling that is not seen in pure $TbMnO_3$. The shift of crystal field excitation frequency can be understood as a consequence of the thermal occupation of the excited crystal field levels of $Tb^{3+}$ and concomitant changes in atomic environment around the $Tb^{3+}$ site. This causes lattice distortions and explain the observed anomalous thermal expansion of the lattice on cooling. [30] We currently have no explanation for the absence of the second electromagnon in the THz spectrum of $TbMnO_3$.

The $E^\omega \perp a$ polarized THz absorption spectra is almost temperature independent above 100 K and below 50 K. However, an abrupt decrease in absorption on cooling from 100 K to 50 K is observed, with change rate that increases with increasing $x$. A similar decrease of absorption was observed (but not explained) by Takahashi et al. in pure $TbMnO_3$. [16] This absorption cannot be ascribed to magnetic excitations above $T_N$, as it was previously explained, and also because this would instead lead to an increase in losses with decreasing temperature below $T_N$. The broad absorption peaking near 50 cm$^{-1}$ in the $E^\omega \parallel b$ polarized absorption spectra above 100 K likely originates from a two-phonon difference process, which expires below 40 K, when the zone-boundary acoustic phonons involved in this process are no longer thermally populated. The spectral component at 80 cm$^{-1}$ is clearly seen at all temperatures in the absorption spectra of both $TbMn_{0.98}Fe_{0.02}O_3$ and $TbMn_{0.96}Fe_{0.04}O_3$, which is assigned to a crystal field excitation. [40] The origin of the band near 12 cm$^{-1}$ in $E^{\omega} \parallel b$ polarized absorption spectra of Fe-substituted compounds is not clear. The existence of this band in both $E^{\omega} \parallel a$ and $E^{\omega} \parallel b$ absorption spectra of TbMn$_{0.96}$Fe$_{0.04}$O$_3$ points for a deep change in the magnetic structure, induced by the Mn$^{3+}$ substitution by Fe$^{3+}$.

The intensity transfer observed in the $E^{\omega} \parallel a$ polarized absorption spectra can be quantitatively described by looking at the low frequency infrared spectra displayed in Figure 9. The lowest frequency band assigned to an optical phonon at ~115 cm$^{-1}$ (marked in Figure 9 by a star) loses intensity on cooling, while the electromagnon and the crystal field excitation gain strength at low temperatures. To quantitatively analyze the intensity transfer, we calculated the spectral weight of these two excitations through the equation: [16]

$$S = \frac{2m_0 V}{\pi e^2} \int_{\omega_1}^{\omega_2} \omega \epsilon''(\omega) d\omega \qquad (6)$$

where $m_0$, $e$ and $V$ are the free electron mass, electron charge and primitive unit cell volume, respectively. The temperature dependence of the spectral weight change relatively to the value at 250 K, $\Delta S = S(T) - S(250\ K)$, is depicted in the inset of Figure 9 for the electromagnon plus crystal field excitation, and the lowest energy polar phonon. The spectral weight was calculated in the frequency ranges $(\omega_1, \omega_2) = (8, 70)$ and $(90, 123)$ cm$^{-1}$ for the electromagnons (region I) and phonon (region II), respectively. The spectral weight of the electromagnon and crystal field excitation considerably increases below 150 K. It is also below that temperature that the lowest-lying optical phonon exhibits a decrease in spectral weight. Indeed, it can be seen from the inset of Figure 9 that the change in spectral weight of the phonon largely accounts for that of the electromagnon and crystal field excitation. This is consistent with what was previously reported in TbMnO$_3$. [16] Moreover, Figure 6 shows that not only the phonon near 115 cm$^{-1}$ transfers its strength to the THz excitation. Also, the phonons near 190, 240 and 400 cm$^{-1}$ reduce their intensities below T$_N$ indicating their coupling with electromagnon and crystal field excitations at 40 and 135 cm$^{-1}$.

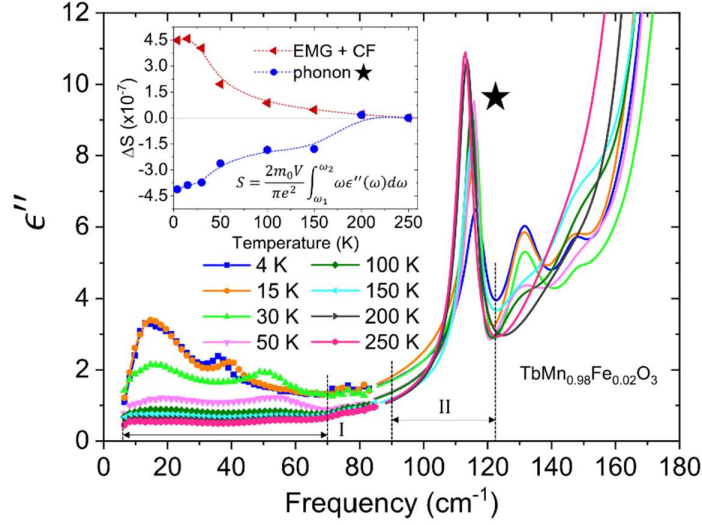

**Figure 9.** Temperature dependence of the $\epsilon''$ spectra for $\boldsymbol{E^\omega} \parallel a$ of TbMn$_{0.98}$Fe$_{0.02}$O$_3$ up to 180 cm$^{-1}$. Inset: Temperature dependence of the change in spectral weight (Equation 6) of the electromagnon (EMG) plus crystal-field (CF) excitation and the lowest-lying optical phonon around 115 cm$^{-1}$ (signaled by a star) from their values at 250 K.

### 4.3. What happens in the paramagnetic phase?

The experimental results here reported evidence interesting changes in the magnetization and lattice distortions, mirrored by both IR and Raman-active phonons, in all compounds in the paramagnetic phase, well above $T_N$. Such observations were reported in TbMnO$_3$, but their explanation is still controversial. Our results show that the Fe-substitution has strong impact on the low temperature magnetic and ferroelectric phases, but not in the temperature dependence of the magnetization in the paramagnetic phase. Anomalous temperature dependence of the magnetization, Raman-active phonons and lattice parameters have been also reported at 150 K in TbFeO$_3$, well within its antiferromagnetic phase ($T_N$ = 650 K). [34] From these observations, we conclude that the changes in both magnetization and crystal lattice are due to the Tb$^{3+}$ and/or O$^{2-}$ atoms. The temperature variations observed in the oscillator strength of the polar optical phonons, in the spectral range where oxygen motions contribute to the lattice vibrations, evidence octahedra deformations involving oxygen atoms in the paramagnetic phase. The anomalous temperature dependence of the lattice parameters, ascertained for both TbMnO$_3$ and TbFeO$_3$, supports the existence of such deformations. Therefore, we propose that oxygen shifts occur around 200 K. The oxygen movements can be explained by the different temperature occupancy of the crystal field electronic Tb$^{3+}$ levels. The oxygen movements explain the changes on the super-exchange integrals and, consequently, the abnormal temperature dependence of magnetization above $T_N$.

The $Mn^{3+}$ substitution by $Fe^{3+}$ has another interesting effect on the coupling between different lattice and electronic excitations in the paramagnetic phase. Regarding the temperature dependence of the Raman-active $Tb^{3+}$ oscillations, significant deviations of the frequency of the Tb(*a*)-oscillations are observed only in Fe-substituted compounds, just below 100 K, while no deviations could be ascertained in $TbMnO_3$. These deviations not only mirror the anisotropy in these compounds but also show a strong interplay between $Tb^{3+}$ and $Fe^{3+}$ spins, as in $TbFeO_3$ [54], even at small concentrations ($x < 0.05$).

## 5. Conclusions

This work evidences the deep influence of the $Mn^{3+}$ substitution by $Fe^{3+}$ in the low temperature magnetic phase sequence, static and dynamic magnetoelectricity, spin-structure and ferroelectricity, as well as, spin-phonon coupling in $TbMn_{1-x}Fe_xO_3$ ($0 \leq x \leq 0.04$). This substitution is found to reinforce the ferromagnetic interactions against the antiferromagnetic ones, leading to a decrease of both magnetic phase transition temperatures and to the spin structure changes that allows for electric polarization. As $Fe^{3+}$ concentration increases up to 4%, the spin structure helicity decreases, and the maximum electric polarization magnitude decreases as a consequence.

The low energy electromagnon excitation is found to be rather strong in the non-ferroelectric and magnetic collinear sinusoidal antiferromagnetic phase in the Fe-substituted compounds, contrarily to $TbMnO_3$. The activation of electromagnon in the spin-collinear phase can be understood in the framework of the exchange-striction mechanism, and corroborates the change of the magnetic structure induced by the presence of $Fe^{3+}$ in the lattice, enabling the activation of electromagnon in the paraelectric and collinear sinusoidal modulated antiferromagnetic phase, stable between $T_C$ e and $T_N$. Traces of the Debye-like background and crystal field absorption remain in the THz spectra up to room temperature. A detailed analysis of IR reflectivity spectra revealed that the spectral weight of the THz excitations in $TbMn_{0.98}Fe_{0.02}O_3$ mainly originates from the lowest-lying optical phonon near 115 cm$^{-1}$, as previously reported for pure $TbMnO_3$. The mode near 135 cm$^{-1}$ is assigned to a crystal field excitation, which receives its dielectric strength from phonons up to 400 cm$^{-1}$. Raman spectra reveal a spin-phonon coupling above $T_N$, in the Fe-substituted compounds.

The abnormal temperature dependence of magnetization around 200 K is explained by changes in the crystal field electronic levels of $Tb^{3+}$ and oxygen shifts. This evidences a magnetoelastic coupling in the studied materials, even in the paramagnetic phase.

**Acknowledgements**

This work was supported by the VEGA 2/0137/19 project, the Czech Science Foundation (Project No. 21-06802S), the MŠMT Project SOLID 21- CZ.02.1.01/0.0/0.0/16_019/0000760, PTDC/FIS-MAC/29454/2017, NORTE/01/0145/FEDER/028538, IFIMUP: Norte-070124-FEDER-000070; NECL: NORTE-01-0145- FEDER-022096, UIDB/04968/2020 and UID/NAN/50024/2019.

**Competing interests**

The authors declare no competing interests.

# Supplemental Information

## Modifying the Magnetoelectric Coupling in TbMnO$_3$ by low-level Fe$^{3+}$ Substitution

A. Maia[1], R. Vilarinho[2], C. Kadlec[1], M. Lebeda[1], M. Mihalik jr.[3], M. Zentková[3], M. Mihalik[3], J. Agostinho Moreira[2], S. Kamba[1]

[1]*Institute of Physics of the Czech Academy of Sciences, Na Slovance 2, 182 21 Prague 8, Czech Republic*

[2]*IFIMUP, Physics and Astronomy Department, Faculty of Sciences, University of Porto, Rua do Campo Alegre 687, s/n- 4169-007 Porto, Portugal*

[3]*Institute of Experimental Physics of the Slovak Academy of Sciences, Watsonova 47, Košice, Slovak Republic*

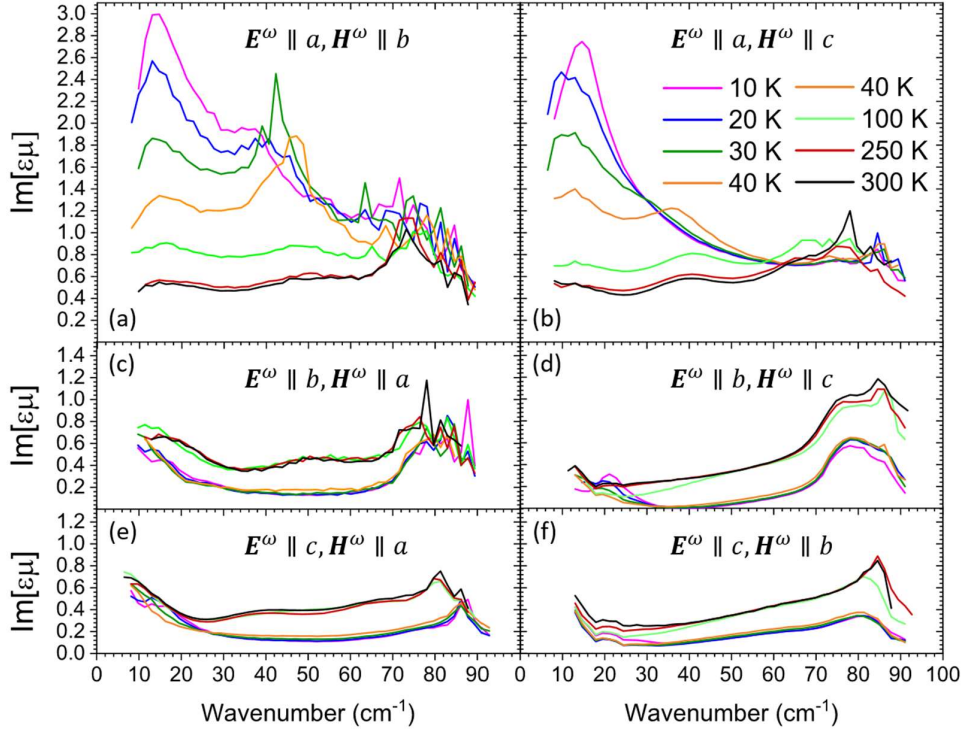

**Figure S1.** Temperature dependence of the $Re[\epsilon\mu]$ and $Im[\epsilon\mu]$ spectra of TbMn$_{0.96}$Fe$_{0.04}$O$_3$ for all the $\boldsymbol{E^\omega}$, $\boldsymbol{H^\omega}$ polarization configurations of the THz radiation. The sharp peak seen at 100 K near 45 cm$^{-1}$ is a consequence of high inaccuracy at this frequency due to high opacity of the sample.

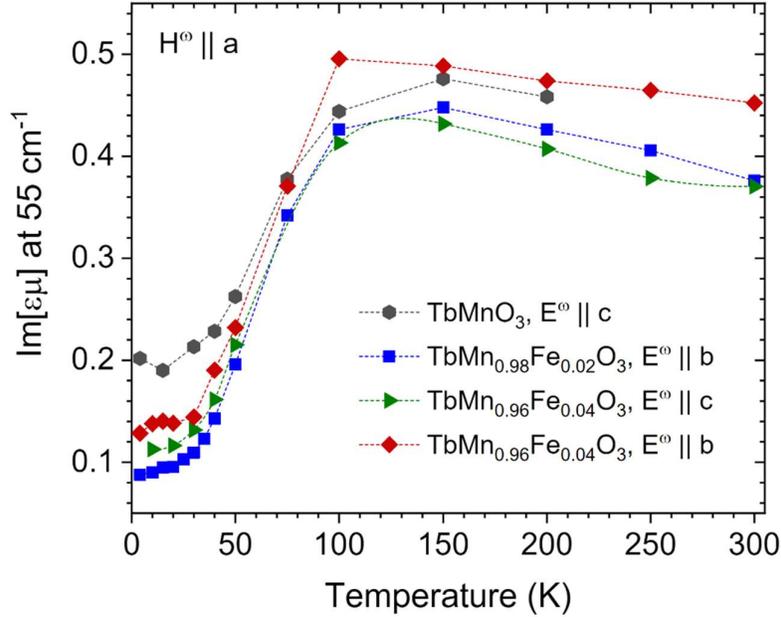

**Figure S2.** Temperature dependence of the $Im[\epsilon\mu]$ spectra at 55 cm$^{-1}$ for $\boldsymbol{H^\omega} \parallel a$ of TbMn$_{1-x}$Fe$_x$O$_3$ for x = 0, 0.02 and 0.04.

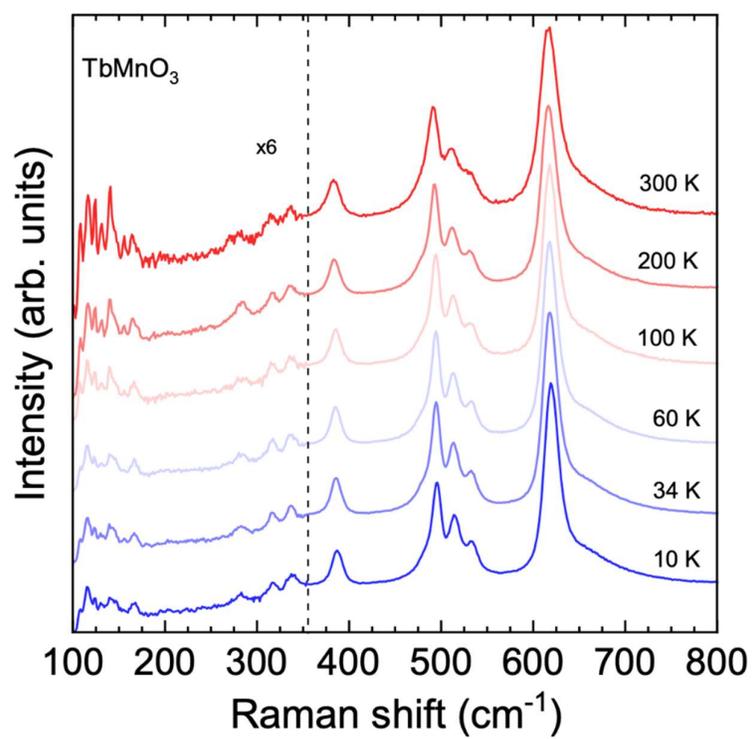

**Figure S3.** Unpolarized Raman spectra of TbMnO$_3$ recorded at several temperatures. The spectra below 354 cm$^{-1}$ are multiplied by 6 for clarity.

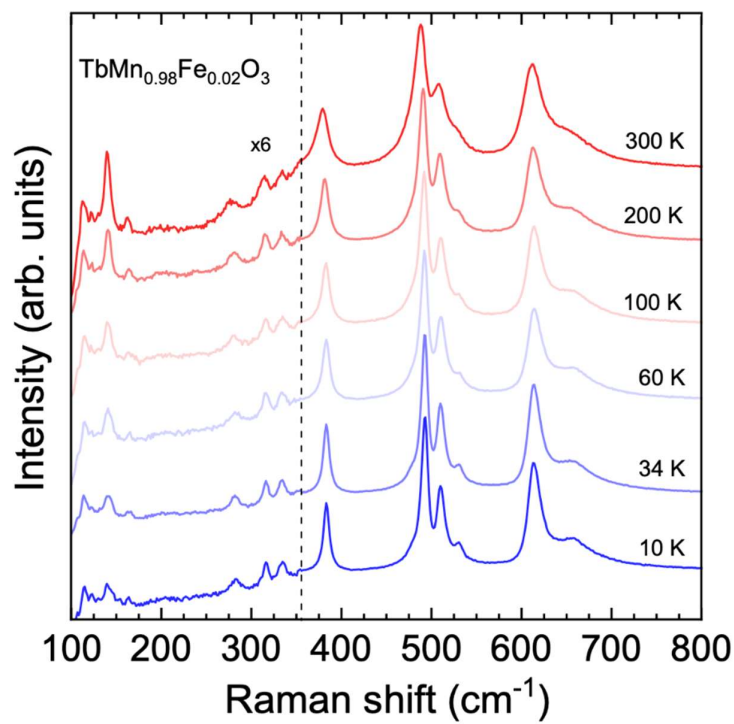

**Figure S4.** Unpolarized Raman spectra of TbMn$_{0.98}$Fe$_{0.02}$O$_3$ recorded at several temperatures. The spectra below 354 cm$^{-1}$ are multiplied by 6 for clarity.

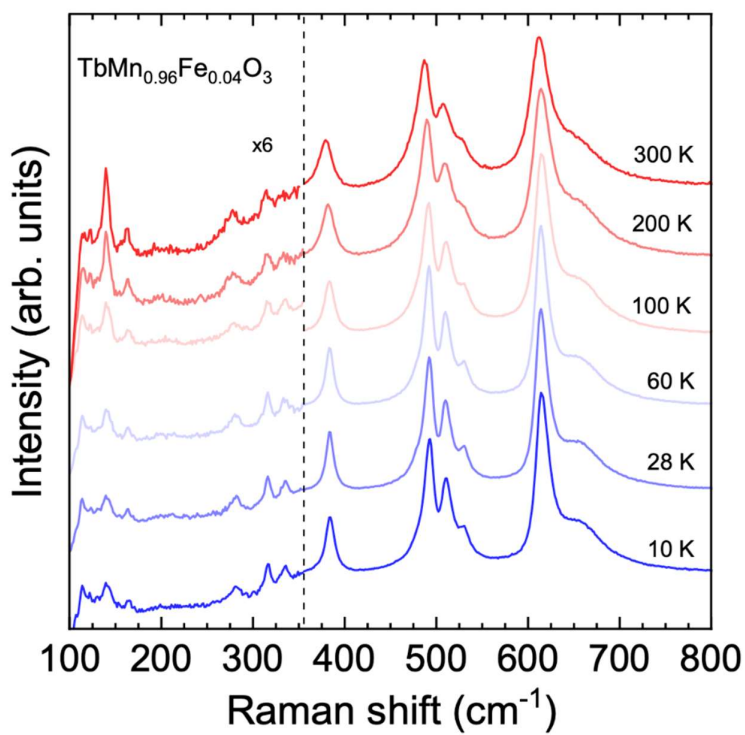

**Figure S5.** Unpolarized Raman spectra of TbMn$_{0.96}$Fe$_{0.04}$O$_3$ recorded at several temperatures. The spectra below 354 cm$^{-1}$ are multiplied by 6 for clarity.